\documentclass[showpacs,pre,twocolumn,amsmath,amssymb,epsfig]{revtex4-1}

\usepackage{graphicx}
\usepackage{psfrag}
\usepackage{epsfig}
\usepackage{dcolumn}
\usepackage{bm}
\newcommand     {\beq}[1]         { \begin{equation} #1 \end{equation} }

\begin{document}

\title{Crackling noise in three-point bending of heterogeneous materials} 

\author{G\'abor Tim\'ar}
\affiliation{Department of Theoretical Physics, University of Debrecen,
P.O. Box 5, H-4010 Debrecen, Hungary}

\author{Ferenc Kun}
\email{feri@dtp.atomki.hu}
\affiliation{Department of Theoretical Physics, University of Debrecen,
P.O. Box 5, H-4010 Debrecen, Hungary}

\begin{abstract}
We study the crackling noise emerging during single crack propagation 
in a specimen under three-point bending conditions. Computer
simulations are carried out in the framework of a discrete element
model where the specimen is discretized in terms of convex polygons and
cohesive elements are represented by beams. 
Computer simulations revealed that
fracture proceeds in bursts whose size and waiting time
distributions have a power law functional form with an exponential
cutoff. Controlling the degree of brittleness of the sample by the
amount of disorder, we obtain a scaling form for the characteristic
quantities of crackling noise of quasi-brittle materials. Analyzing
the spatial structure of damage we show that ahead of the crack tip a
process zone is formed as a random sequence of broken and intact
mesoscopic elements. We characterize the statistics of the shrinking
and expanding steps of the process zone and determine the damage
profile in the vicinity of the crack tip. 
\end{abstract}
\pacs{89.75.Da, 46.50.+a, 05.90.+m}
\maketitle

\section{Introduction}
The brittle fracture of materials has two substantially different
scenarios depending on the amount of structural disorder: for
homogeneous materials such as crystalline solids at the critical
stress a single crack is formed which propagates in an unstable
manner. However, in materials with a high degree of heterogeneity
fracture develops progressively, i.e.\ under an increasing external
load first microcracks nucleate at local weaknesses which may then
undergo several steps of growth and arrest. Finally macroscopic
fracture occurs as the culmination of the gradual accumulation of
damage \cite{johansen_critical_2000,alava_statistical_2006}. The
nucleation and growth of cracks is accompanied by the 
emission of elastic waves which can be recorded in the form of
acoustic noise \cite{carpinteri_acoustic_book}. Measuring acoustic
emissions (AE) on loaded specimens 
is the primary source of information on the microscopic dynamics of the
fracture of heterogeneous brittle materials
\cite{alava_statistical_2006,petri_experimental_1994,deschanel_experimental_2009}.
During the last two decades a large number of acoustic emission
experiments were carried out on different materials under
quasi-statically increasing and constant external loads. These
experiments revealed that the energy of acoustic bursts and the
waiting times between consecutive events are characterized by power
law distributions. The value of the exponents of crackling noise is
found to be characteristic for the type of fracture, i.e.\ for ductile
fracture the exponents are larger than for brittle breaking since
large acoustic bursts are suppressed in ductile materials \cite{kun_structure_2004,deschanel_experimental_2009}. 

It has long been
recognized that in spite of the widely different length scales,
acoustic emissions of fracturing solids and earthquakes share several
common features \cite{carpinteri_acoustic_book}. Recently, it has been
pointed out that the 
probability distribution of waiting times $T$ between
consecutive earthquakes can be described by a simple scaling form 
\beq{
P(T) \sim Rf(RT),
\label{eq:corral}
}
where $R$ denotes the mean rate of events in the time window
considered. The generic law Eq.\ (\ref{eq:corral}) proved to be valid
for all geographical regions unless the time window is sufficiently
broad to fulfill the condition of stationarity
\cite{corral_long-term_2004}.  Laboratory experiments on the fracture
of different types of materials have revealed that the scaling form
Eq.\ (\ref{eq:corral}) is also valid for acoustic emissions, even the
scaling function $f(x)$ proved to have the same functional form, i.e.\
a $gamma$ distribution is obtained $f(x)=Ax^{\gamma}exp{(-x/B)}$ which
means a power law decay
followed by an exponential cutoff \cite{davidsen_scaling_2007}. 

Very
recently the statistical features of acoustic emissions have been
analyzed during three-point bending tests of notched concrete
specimens. The notch ensures that the formation of microcracks is
dominated by strong spatial correlations in a narrow cross section of
the specimen. Varying the detection threshold of acoustic signals, for
the waiting time distributions the scaling behavior Eq.\
(\ref{eq:corral}) was recovered
\cite{niccolini_crackling_2009,niccolini_self-similarity_2009}. In
Ref.\ \cite{niccolini_self-similarity_2009} the three-point bending
experiment was modelled by discretizing the bar in terms of a bundle
of fibers. It was a crucial feature of the model that after fiber
breakings the load was redistributed equally in the bundle, i.e.\
spatial correlation of microfractures were not taken into
account. Computer simulation of the bending process revealed the same
scaling structure Eq.\ 
(\ref{eq:corral}) of the numerical results, however, for the scaling
function $f$ an exponential form was obtained $f\sim
\exp{(-RT)}$. These experimental and theoretical results demonstrate
the importance of the range of stress redistribution and correlations of
consecutive acoustic events in the statistics of microfractures. 

Motivated by these theoretical and experimental findings, in the
present paper we investigate the fracture of heterogeneous materials
under three-point bending conditions by means of a discrete element
model (DEM). Our two-dimensional DEM approach provides a realistic
representation of the microstructure of the material, the formation of
microcracks, and the emerging complicated stress field naturally
accounting for the correlation of microfractures. We analyze both the
temporal evolution and spatial structure of damage varying the degree
of heterogeneity of the material. Our investigations showed that the
fracture proceeds in bursts whose size and waiting time
distributions have power law behavior with an exponential
cutoff. Varying the amount of disorder of the sample we can control
the degree of brittleness of the final failure of the material. A
scaling form is determined in terms of which the distributions
obtained at different amounts of disorder can be collapsed on a master
curve. Ahead of the crack tip a process zone is formed composed of
broken and intact mesoscopic elements. Our DEM approach allows us to
carry out a detailed analysis of the spatial structure of damage as
well. 

\section{Discrete element model for heterogeneous materials}
Three-point bending is a standard engineering test where a bar shaped
specimen is clamped at the two ends and a point load is applied in the
middle perpendicular to the longer axis of the bar. Under an
increasing load the bar bends and finally breaks due to a crack which
appears in the middle along the load direction. This testing method is
mainly used in the engineering literature to characterize the
quasi-static fracture strength of structural materials such as
concrete \cite{mier_fracture_1997}. On the
other hand, three-point bending experiments provide an excellent
opportunity to study the propagation of a single crack in a disordered
environment which is a challenging problem for the statistical physics
of fracture \cite{alava_statistical_2006}.

\begin{figure}
\begin{center}
\epsfig{ bbllx=0,bblly=0,bburx=570,bbury=215,file=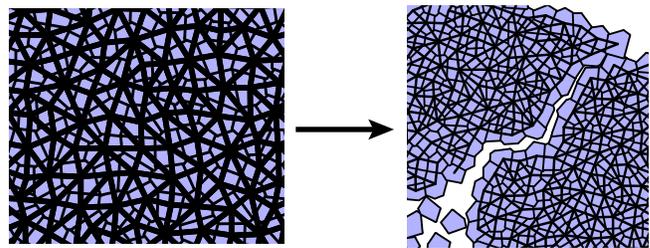,
width=8.5cm}
\end{center}
\caption{ {\it (Color online)}
$(left)$ Neighboring polygons of the initial Voronoi
tessellation are connected by beams. This way a triangular beam
lattice is obtained. $(right)$ Due to subsequent breaking of beams a
crack forms along the edge of polygons. 
}
\label{fig:polygon}
\end{figure}

Recently, we have worked out a two-dimensional dynamical model of
deformable, breakable granular solids, which enables us to perform
molecular dynamics simulation of
fracture and fragmentation of solids in various experimental
situations \cite{kun_study_1996,behera_fragmentation_2005,daddetta_application_2002,kun_transitiondamage_1999}. Our model is an extension of
those models which are used
to study the behavior of granular materials applying randomly shaped
convex polygons to describe grains \cite{kun_study_1996}. The initial set of
polygons is obtained by the Voronoi tessellation of a rectangle from
which specimens of appropriate shapes can be cut out. The average
polygon 
size $l_p$ sets the characteristic length scale of the model
system. The polygons are 
considered to be rigid bodies which can overlap when pressed against
each other. We 
introduce a repulsive force between the overlapping particles
proportional to the overlap area
\cite{kun_study_1996,behera_fragmentation_2005,daddetta_application_2002,kun_transitiondamage_1999}.
To capture the
elastic behavior of solids we connect the
unbreakable, undeformable polygons (grains) by elastic beams. The
beams have two important roles in the model construction: they ensure
cohesion and they are able to break which is essential to model
fracture processes. The beams can be elongated, compressed, sheared
and bent so that they exert forces and torques on the polygons to
which they are attached. Figure \ref{fig:polygon} presents an example
of the polygon structure and the beam lattice attached to the polygons.

\begin{figure}
\begin{center}
\epsfig{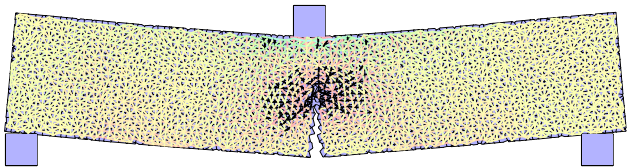}
\end{center}
\caption{{\it (Color online)}
Three-point bending of a bar composed of polygonal
particles. The particles are coupled by elastic beams which are
colored according to the longitudinal deformation (yellow: nearly
unstressed beams; red and black:
elongated beams; blue and green: compressed beams). Beams are allowed
to break solely along the center line of the bar. 
A relatively small sample is presented to have a clear view
on the details of the model construction. The two loading plates at
the bottom are fixed while the third one on the top moves downward.
}
\label{fig:sample}
\end{figure}

In the simulations a bar shaped specimen is considered with longer 
and shorter side lengths $L$ and $L_c$, respectively. 
In order to 
make a realistic representation of three-point loading, 
the three loading plates are realized by additional polygonal
elements, i.e.\ squares in Fig.\ \ref{fig:sample} with side
length $S=5l_p$ much smaller than the longer side $L=200l_p$ of the bar
$S\ll L$. 
These loading plates interact with the particles of the
bar via the overlap force, however, no beams are coupled to
them. Strain controlled loading of the bar is implemented in such a
way that the two loading plates at the bottom are fixed while the
third one on the top is moved vertically downward in Fig.\
\ref{fig:sample} with a constant speed $v_0$. The moving plate
overlaps the boundary polygons on the top of the bar which results in
an increasing loading force. The stiffness of the plates is set high
enough to keep the overlap below $20\%$ of the average polygon area. 
Simulations were carried out varying the value
of $v_0$ in a range, which allows for an efficient
damping of the elastic waves and ensures a reasonable CPU time for the
computations. The main advantage of three-point bending tests is that
the highly stressed zone, where the crack appears, falls in the middle
of the bar which helps to make efficient monitoring of the fracture
process. In order to simplify the numerical measurements on crack
propagation, we introduce a ``weak'' line in the middle of the bar in
such a way that solely those beams are allowed to break which connect
the two sides of the line (see Fig.\ \ref{fig:sample}). 

\subsection{Disordered beam breaking}
The beams, modeling cohesive forces between grains, can be broken
according to a physical breaking
rule, which takes into account the stretching and bending of the
connections \cite{behera_fragmentation_2005,daddetta_application_2002}
\begin{eqnarray}
\left(\frac{\varepsilon}{\varepsilon_{th}}\right)^2
+\frac{max(|\Theta_1|, |\Theta_2|)}{\Theta_{th}} \geq 1.
\label{eq:break}
\end{eqnarray}
Here $\varepsilon$ denotes the longitudinal deformation of a beam,
while $\Theta_1$ and $\Theta_2$ are bending angles at the two beam ends.
The breaking rule Eq.\ (\ref{eq:break}) contains two parameters
$\varepsilon_{th}, \Theta_{th}$ controlling the relative importance of
the 
stretching and bending breaking modes, respectively.  The energy
stored in a beam just before breaking is released in the breakage
giving rise to energy dissipation.
 At the broken beams along the surface of the polygons
cracks are generated inside the solid and as a result of the successive
beam breaking the solid falls apart (see Fig.\ \ref{fig:polygon}). 
The time evolution
of the polygonal solid is obtained by
solving the equations of motion of the individual polygons. At each
iteration step we evaluate the breaking criterion Eq.\
(\ref{eq:break}) and remove those beams which fulfill the condition.
The simulation is continued until all beams break along the weak
interface and the specimen falls apart. 
(For more details of the model construction see
Refs.\ \cite{kun_study_1996,behera_fragmentation_2005}.)

The breaking parameters $\varepsilon_{th}$ and $\Theta_{th}$ of beams
are stochastic variables in the model, i.e.\ they are sampled from
probability density functions $p(\varepsilon_{th})$ and
$p(\Theta_{th})$. The Weibull distribution provides a
comprehensive description of the stochastic fracture strength of
materials, hence, for both threshold values the Weibull form is
prescribed 
\begin{eqnarray}
p_{\lambda,m}(x)=
\frac{m}{\lambda}\left(\frac{x}{\lambda}\right)^{m-1}e^{-(x/\lambda)^m},
\label{eq:weibull}
\end{eqnarray}
where $x$ denotes the two breaking thresholds $\varepsilon_{th},
\Theta_{th}$. The Weibull distribution has two
parameters: $\lambda$ sets the characteristic scale of threshold
values while the exponent $m$ determines the scatter of the
variable. Increasing the value of the exponent $m$ the width of the
Weibull distribution Eq.\ (\ref{eq:weibull}) decreases and converges
to the delta function in
the limit $m\to \infty$. Varying the scale parameters
$\lambda_{\varepsilon}$ and $\lambda_{\Theta}$ of the breaking
thresholds the relative importance
of stretching and bending can be controlled in the breaking process.
For clarity in the present paper we only investigate the two limiting
cases of beam breaking dominated by pure stretching or bending
with the parameter settings $\lambda_{\varepsilon}=0.05$,
$\lambda_{\Theta}=100$, or $\lambda_{\varepsilon}=100$,
$\lambda_{\Theta}=1$, respectively. The Weibull exponents were changed
in the range $1\leq m \leq 50$ for both threshold distributions in
order to control the amount of disorder in the system. We
note that the elastic constants of beam elements depend on the Young's
modulus of beams, furthermore, also on their length and cross
section. In the model the geometry of beams is determined by the
Voronoi tessellation, i.e.\ the length and cross section of beams are
defined as the distance between the center of mass and the length of
the common side of the two neighboring polygons, respectively. It has
the consequence that besides the strength disorder of beams there is
also structural disorder in the system determined by the initial
Voronoi tessellation.  

\section{Macroscopic response}
\label{sec:macro}
In our strain controlled three-point bending experiment the mechanical
response of the material can be characterized numerically by measuring
the force $F$ acting on the moving plate at the top of the sample as a
function of time $t$ (see Fig.\ \ref{fig:sample}). 
\begin{figure} [h]
\begin{center}
\epsfig{bbllx=10,bblly=507,bburx=350,bbury=792,
file=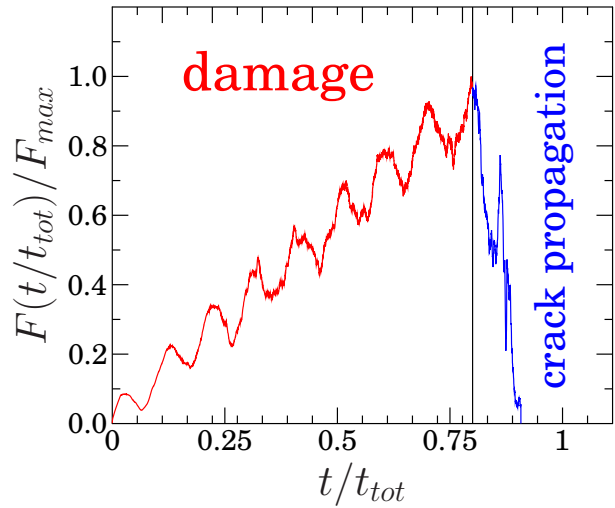, width=9.5cm}  
\caption{{\it (Color online)} Force $F$ normalized by the maximum
  value $F_{max}$ as a function of time $t$ during the
  loading process. Since the deflection of the bar is
  proportional to $t$ the curve can be considered as the constitutive
  curve of the sample. Oscillations occur due to elastic waves
  generated by the loading process. After the peak the decreasing part
  of $F(t)$ indicates stable crack propagation where our
  analysis is focused. $t_{tot}$ denotes the time of the last beam breaking.
}
\label{fig:fcurve}
\end{center}
\end{figure}
Figure \ref{fig:fcurve} presents the force-time curve obtained for a
sample with the Weibull exponent $m=2.0$ in the stretching
limit. Since the loading plate moves at a constant speed the
deflection of the bar is proportional to $t$ so that $F(t)$ can be
considered to be the 
constitutive curve of the sample.
It can be seen that the macroscopic response is linear
all the way up to the peak, where the force drops 
suddenly. This drop becomes more and more drastic as we increase the
brittleness of the sample by increasing the value of the Weibull
exponent.  
At the beginning of the loading process, the smooth oscillations about
the linear in the constitutive curve arise due to elastic waves
generated by the loading plate. As the force approaches its maximum, the
curve becomes more and more noisy due to internal damage being
accumulated in the form of microcracks nucleating throughout the
breakable interface. It can be observed in Fig.\ \ref{fig:fcurve} that
after the maximum, the force drops rather drastically, 
however, the failure is not totally abrupt. The sharp drop-down of the
force is followed by stable crack propagation where the crack
gradually advances until the sample falls apart. 

As the bar is loaded at a constant strain-rate, microcracks -
corresponding to uncorrelated beam breaking events - start nucleating
throughout the interface. This way damage is accumulated inside the
sample before the onset time of crack propagation. Local beam breaking
inside the sample is always brittle, however, the disorder of breaking
thresholds can result in a quasi-brittle macroscopic response where
the constitutive curve exhibits a non-linear behavior. Due to the
disturbing effect of elastic waves, numerically it is difficult to
quantify the strength of non-linearity of the $F(t)$ curve in Fig.\
\ref{fig:fcurve}. Hence, we characterize the degree of brittleness of
the sample and its dependence on the amount of threshold disorder by
measuring the accumulated damage prior to the peak of the force as a
function of the Weibull exponent $m$.  
\begin{figure}[!ht]
\begin{center}
\epsfig{bbllx=1,bblly=507,bburx=341,bbury=792,
file=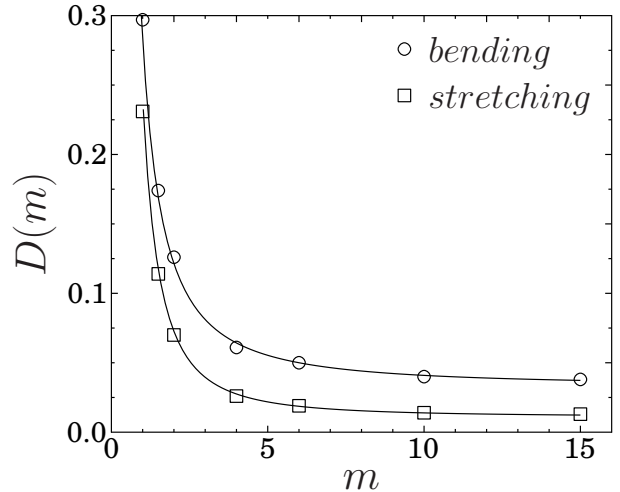, width=9.5cm}  
\caption{Damage accumulated up to the peak of
  $F(t)$ as a function of 
  the Weibull exponent $m$. The curves can be very well fitted with the
  functional form Eq.\ (\ref{damrat}). The value of the exponent is
  $\mu = 1.9$ and $\mu = 1.5$ for the stretching and bending limits,
  respectively. }
\label{fig:damratio}
\end{center}
\end{figure}
Figure \ref{fig:damratio} shows the damage parameter $D$ of the model,
defined as the fraction of beams broken
before a single crack starts propagating, for the stretching and
bending limits, as a function of the Weibull exponent $m$. It can be
observed that the curves can be very well fitted with the functional
form 
\begin{equation}
D(m) = B + A m^{-\mu},
\label{damrat}
\end{equation}
where all parameters $A$, $B$ and $\mu$ proved to be different 
for the stretching
and bending limits. The power law form of $D(m)$ can be
motivated by the following simplified assumption: 
Let us consider a mean-field approximation of the system, where all
the beams along the interface share the same $\varepsilon$ strain at any
given time during the process.
This way the breakable interface of the bar is substituted by a
parallel bundle of beams with equal load sharing, whose breaking
process can easily be described analytically
\cite{kun_damage_2000-1,hidalgo_bursts_2001,kun_damage_2008}. The
fraction of intact 
beams at any $\varepsilon$ can be given as $1 - P(\varepsilon)$ where
$P(\varepsilon)$ is the cumulative probability distribution of the
breaking thresholds. The macroscopic stress $\sigma$ as a function of
strain $\varepsilon$ can then be written in the form 
\begin{eqnarray}
\sigma(\varepsilon) = \left[1 -
P(\varepsilon)\right] E \varepsilon = e^{-(\varepsilon / \lambda)^m} E
\varepsilon,
\label{eq:fbm_1}
\end{eqnarray}
where $E$ is the Young's modulus of the beams. Under strain controlled
loading of the bundle, stable crack propagation starts at the peak of
the constitutive curve $\sigma(\varepsilon)$. After differentiating
Eq.\ (\ref{eq:fbm_1}) the position of the maximum $\varepsilon_c$
reads as $\varepsilon_c = \lambda (1 / m)^{1/m}$ for the Weibull
distribution. The fraction of broken beams accumulated up to the peak
of $\sigma(\varepsilon)$ can be obtained by plugging $\varepsilon_c$
into the cumulative distribution of thresholds $P(\varepsilon_c)$,
hence, the damage parameter $D$ as a function of the Weibull exponent
$m$ can be cast into the final form for large enough $m$ values 
\begin{eqnarray}
D(m) \approx 1 - e^{-(1 / m)} \sim m^{-1}.
\end{eqnarray}
The numerical results on the amount of damage prior to the force peak
in Fig.\ \ref{fig:damratio} are consistent with the above analytic
prediction. The higher value of the measured exponents $\mu^s=1.9\pm 0.1$
(stretching) and $\mu^b=1.5\pm 0.1$ (bending) is the 
consequence of the strain gradient in the load direction, which was
completely neglected in the analytic calculations. Note that in the
limit of high $m$ values, 
the amount of damage does not converge to zero, instead it takes a
finite value $B>0$. The non-zero value of $B$ in Eq.\ 
(\ref{damrat}) can be attributed to the structural disorder in the
sample, which is present and is the same for all values of the Weibull
exponent. This structural disorder gives rise to fluctuations of the 
beam parameters which in turn result in a noisy breaking sequence in
spite of the constant breaking parameters
\cite{behera_fragmentation_2005,daddetta_application_2002}.  

Perfectly brittle failure of the bar would be characterized by a
linear behavior of $F(t)$ up to the maximum without any damaging
which is then followed by an abrupt breaking. Our simulation results
demonstrate that varying the amount of threshold disorder we can
control the degree of brittleness of the DEM sample from highly (but
not perfectly) brittle to quasi-brittle. It is a very interesting
question how the degree of brittleness affects the properties of
crackling noise and the spatial structure of damage along the interface.

\section{Crackling noise during crack propagation}
A snapshot of the computer simulation of a three-point bending test
is presented in Figure \ref{fig:sample}. The color code shows
that the bottom of the specimen is highly elongated that's why the
crack starts here. In the vicinity of the crack tip the beams are
strongly elongated indicating a high stress concentration ahead of the
crack which provides the driving force for crack propagation. At the
top of the bar the color code indicates that compressive 
stresses arise. 
\begin{figure}
\begin{center}
\epsfig{bbllx=10,bblly=510,bburx=310,bbury=770,file=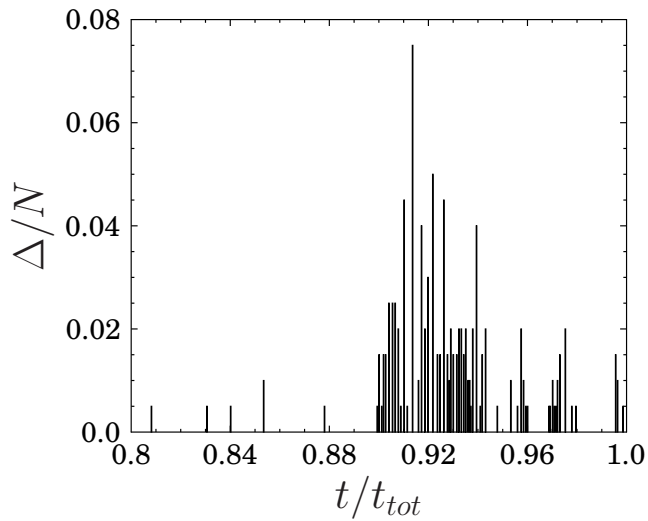,
  width=8.3cm}
 \caption{
Time series of bursts in a single fracture simulation. The bursts are
correlated breaking sequences of beams which then result in sudden
jumps of the extending crack. $N$ denotes the total number of beams along
the weak interface where the crack propagates. For all the simulations
its value was set to $N=200$. At the beginning of the loading process,
for a considerable time no breaking occurs, most of the breaking
events appear at larger deflections beyond the peak of the
constitutive curve (see Fig.\ \ref{fig:fcurve}). Hence,
we magnify the final section of the bending process. 
}
\label{fig:idosor}
\end{center}
\end{figure}
The constant speed of the loading plate ensures a strain controlled
loading of the specimen at a fixed strain rate. The low value of the
loading speed
has the consequence that in each iteration step of the molecular
dynamics simulation either no beam breaking occurs or only  a single
beam breaks. After a local breaking event the stress gets
redistributed increasing the stress concentration on the intact
elements ahead of the crack. This load redistribution may give rise to
additional 
breakings resulting in a correlated trail of breaking events. 
\begin{figure}
\begin{center}
\epsfig{bbllx=20,bblly=520,bburx=320,bbury=770,
file=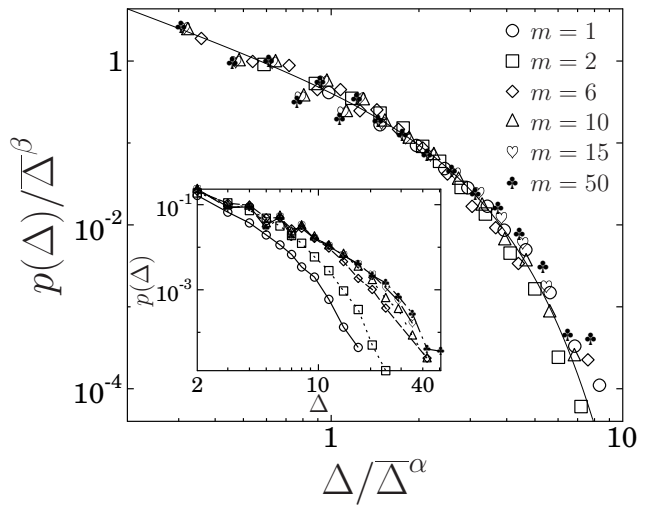, width=8.3cm}  
\caption{Inset: Avalanche size distributions for the absolute
  stretching limit varying the value of the Weibull exponent
  $m$. The main panel presents the excellent data collapse obtained by
  rescaling the distributions with the average burst size according to
  Eq.\ (\ref{eq:scaling}). Scaling exponents: $\alpha^s_{\Delta} = 1.4 \pm 0.5,
  \beta^s_{\Delta} = 1.8 \pm 1$. The parameter values obtained by fitting Eq.\
  (\ref{eq:avaldist})  are $a^s_{\Delta} = 0.55, \tau^s_{\Delta} = 1.3
  \pm 0.2, b^s_{\Delta} = 2.2, \delta^s_{\Delta} = 1.5 \pm 0.3$. 
}
\label{fig:size_s}
\end{center}
\end{figure}

In order
to identify bursts of local breakings we introduce a correlation time
$t_{corr}$: if the time difference of two consecutive beam breakings
occurring at times $t_i$ and $t_{i+1}$ is smaller than the correlation
time $t_{i+1}-t_i < t_{corr}$ the two breakings are considered to belong
to the same burst. The value of the correlation time was chosen in
such a way that it is larger than the time step $\Delta t$ used
in the integration of the equation of motion but it is much smaller
than the total duration $t_{tot}$ of the breaking process, i.e.\ we set
$t_{corr}=10 \Delta t$ for which $10^5 t_{corr}< t_{tot}$ holds.  
The size of bursts $\Delta$ is defined as the
number of beams breaking during the correlated sequence. When
the amount of disorder is very high $m\to 1$, especially in the
bending limit of breakings, it may happen in
DEM simulations that very distant beams break within the correlation
time, however, without any correlation. To obtain information on the
strength of spatial correlations in an avalanche, we calculate the
distance $h_j=|y_j-y_{j+1}|$ between consecutive beam breakings with
the positions $y_j$ and $y_{j+1}$ and sum it up inside an avalanche
$h=\sum_{j=1}^{\Delta-1}h_j$. For a strongly correlated avalanche
where each consecutive breaking occurs on adjacent beams the ratio of
$h$ and of the burst size $\Delta$ is close to the characteristic
polygone size $h/\Delta \approx l_p$. In order to filter out
avalanches dominated by random coincidences we introduce a threshold
value for this ratio, i.e.\ those avalanches for which $h/\Delta > 2
l_p$ holds are removed from the statistics. Computer simulations showed that
in the stretching limit the above condition has no effect, however, in
the bending limit where a high amount of distributed cracking occurs,
about $10\%$ of the avalanches are filtered out due to random
coincidences (compare also to Fig.\ \ref{fig:damratio}).

Figure \ref{fig:idosor} presents the size of bursts in a single
fracture simulation at the time of their appearance. 
It can be seen in the
figure that the bursts are separated by silent periods with variable
length. These waiting times between bursts characterize the duration
of states where the crack tip is pinned due to the presence of some
strong beams. At the beginning of the loading process the bursts are
small compared to the cross section of the specimen (maximum crack
length), however, with increasing deflection of the bar the burst
size $\Delta$ increases and reaches a maximum somewhat before the last
breaking. After the maximum the burst size decreases showing that as
the crack approaches the top of the bar it slows down due to the
compressive zone.

\begin{figure}
\begin{center}
\epsfig{bbllx=20,bblly=520,bburx=310,bbury=770,
file=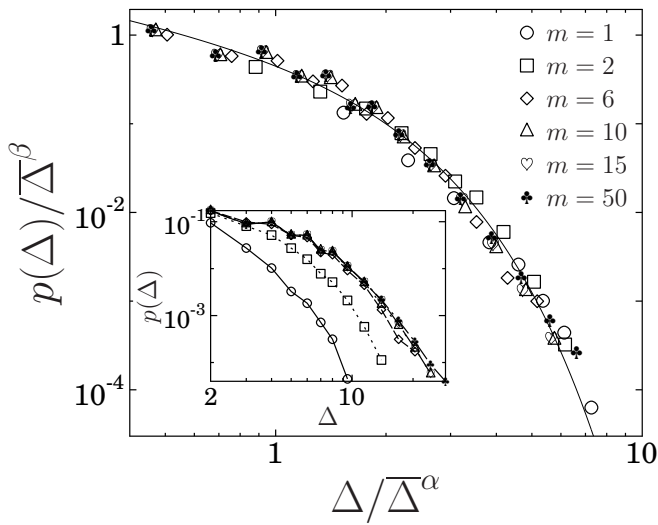, width=8.3cm}  
\caption{ Inset: avalanche size distributions for the absolute bending
  limit. The main panel shows that rescaling the distributions
  according to Eq.\ (\ref{eq:scaling}) an excellent data collapse is
  obtained. Scaling exponents: $\alpha_{\Delta}^b = 1.4 \pm 0.5,
  \beta_{\Delta}^b = 1.8 
  \pm 1$. The fit parameters of the scaling function are $a_{\Delta}^b
  = 0.85, \tau_{\Delta}^b = 0.8 \pm 0.3, b_{\Delta}^b = 1.4,
  \delta_{\Delta}^b = 1.3 \pm 0.3$.  
}
\label{fig:size_b}
\end{center}
\end{figure}

We determined numerically the size distribution of bursts $P(\Delta)$
varying the amount
of disorder in the failure thresholds. The size distribution obtained
at different values of the Weibull exponent is presented in 
the insets of Fig.\ \ref{fig:size_s} and Fig. \ref{fig:size_b} for the
absolute stretching and bending limits, respectively. It is
interesting to note that increasing the Weibull exponent $m$, i.e.\
decreasing the amount of 
disorder, the bursts get larger but the functional form does not
change. 
For small bursts a power law behavior is obtained followed by
a rapidly decreasing cutoff regime. The main panels of Fig.\
\ref{fig:size_s} and Fig.\ \ref{fig:size_b} 
demonstrate that using the average burst size $\overline{\Delta}$ as
a scaling variable, the burst size distributions $P(\Delta)$ obtained
at different $m$ values can be collapsed on a master curve. The data
collapse implies the scaling structure
\begin{eqnarray}
P(\Delta)=\overline{\Delta}^{\beta}
f(\Delta/\overline{\Delta}^{\alpha}),
\label{eq:scaling}
\end{eqnarray}
where the values of the exponents were determined numerically 
$\alpha^s_{\Delta}=1.4\pm 0.5$, $\beta^s_{\Delta}=1.8\pm 1$, and
$\alpha^b_{\Delta}=1.4\pm 
0.5$, $\beta^b_{\Delta}=1.8\pm 1$ which provide the 
best quality collapse for stretching and bending, respectively. 
The scaling function $f$ can be very well fitted by the form
\begin{equation}
f(x)=a x^{-\tau} \displaystyle{e^{-(x/b)^{\delta}}},
\label{eq:avaldist}
\end{equation}
where the parameter values providing the best fit are $a_{\Delta}^s = 0.55, \tau_{\Delta}^s
= 1.3 \pm 0.2, b_{\Delta}^s = 2.2, \delta_{\Delta}^s = 1.5 \pm 0.3$
(stretching), and 
$a_{\Delta}^b = 0.85, \tau_{\Delta}^b = 0.8 \pm 0.1, b_{\Delta}^b =
1.4, \delta_{\Delta}^b = 1.3 \pm 0.3$ (bending). 
The results demonstrate that the growth of the crack is not a smooth
process, the slow driving results in a jerky crack propagation which
is composed of a large number of discrete steps. The growth steps are
sudden outbreaks with a variable length. The correlation of
consecutive local breakings leads to a power law functional form
limited by an exponential cutoff. The
most interesting outcome of the calculations is that the amount of
disorder only affects the characteristic scale of bursts but the
functional form and the value of the power law exponents
$\tau^s_{\Delta}$ and $\tau^b_{\Delta}$ remains the
same. We note the differences between the limits of stretching and
bending dominated breaking, especially the deviation of the exponents
$\tau^s_{\Delta}$ and $\tau^b_{\Delta}$ is beyond the error bars.
The functional form Eq.\ (\ref{eq:avaldist}) has also been found to
provide a good quality description of the amplitude distribution of 
acoustic bursts in three-point bending experiments on concrete
samples \cite{niccolini_crackling_2009,niccolini_self-similarity_2009}. 
\begin{figure}
\begin{center}
\epsfig{bbllx=20,bblly=510,bburx=310,bbury=760,
file=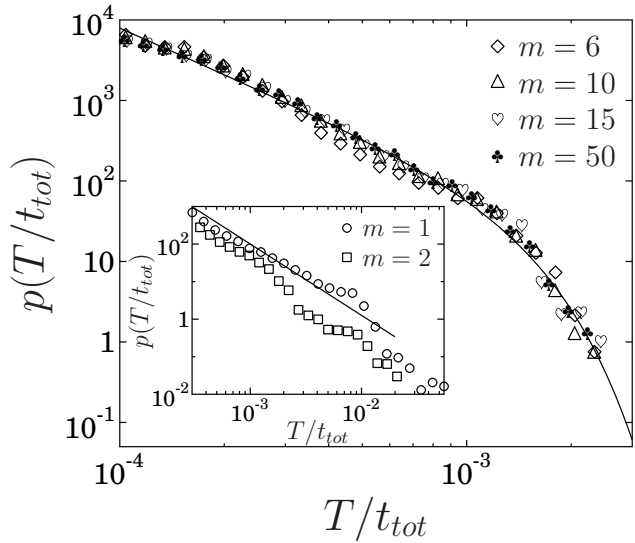, width=8.3cm}  
\caption{Waiting time distributions for absolute stretching. 
The main panel presents curves for low disorder, where the fit was
obtained with the exponent $\tau_T^s=1.9 \pm 0.15$. The inset shows the
corresponding curves for high disorder, where a crossover is obtained
to a lower exponent $\tau_T^s=1.5 \pm 0.1$.
}
\label{fig:wt_s_ld}
\end{center}
\end{figure}

\begin{figure}
\begin{center}
\epsfig{bbllx=1,bblly=507,bburx=341,bbury=792,
file=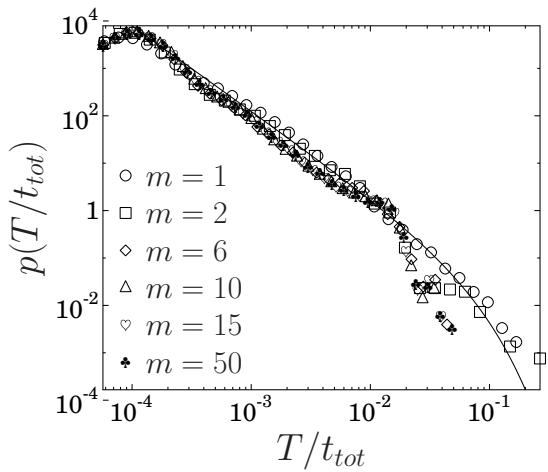, width=8.3cm}  
\caption{Waiting time distributions for the absolute bending 
  limit. The amount of disorder only affects the cutoff but the
  exponent is constant $\tau_T^b=1.8 \pm 0.15$. 
}
\label{fig:wt_b}
\end{center}
\end{figure}

It can be observed in Fig.\ \ref{fig:idosor} that the bursts are
separated by silent periods where no beam breaking occurs. 
The advancing loading plate gradually increases the load on the
system which reactivates the crack after some waiting time $T$. It can be
seen in Fig.\ \ref{fig:idosor} that the 
duration $T$ of these waiting times can vary in a broad range. In
Figure \ref{fig:wt_s_ld} the waiting time
distributions $P(T)$ are presented for the stretching limit
separated for high (inset) and low disorder (main panel). It is
interesting to note that for low enough disorder (main panel of Fig.\
\ref{fig:wt_s_ld}) the distributions $P(T)$ are all the same, no
dependence on the Weibull exponent could be pointed out. The
functional form of $P(T)$ can be very well fitted by the expression 
Eq.\ (\ref{eq:avaldist}) where the value of the exponent
$\tau_T^s=1.9 \pm 0.15$ was obtained. The relatively high value of $\tau_T^s$
implies that long waiting times are very rare in the trail of bursts
when the material is very brittle. 
However, in
the limit of high disorder $m\to 1$ (see inset of Fig.\
\ref{fig:wt_s_ld}) waiting times span a broader range and reach an
order of magnitude larger values than for the very brittle materials
with low disorder. The most remarkable feature of waiting time
distributions is that increasing the disorder the exponent of the
power law regime changes to the lower value $\tau_T^s=1.5$ coinciding
with the recurrence time exponent of one-dimensional random walks.
In the absolute bending limit (see Fig.\ \ref{fig:wt_b}) $P(T)$ has
qualitatively the same behavior as in the stretching limit. Due to
the fragility of the system at all Weibull exponents $m$, the change
of disorder only results in a change of the cutoff, however, the value
of the exponent of the power law regime is constant $\tau_T^b=1.8 \pm 0.15$.

\section{Spatial structure of damage}
It has been discussed in Sec.\ \ref{sec:macro} that at the time of the
force drop in Fig.\ \ref{fig:fcurve}, a crack initiates along the
breakable interface of the sample and proceeds in a jerky way.
The crackling noise analyzed in the previous section characterizes the
temporal fluctuations of the advancing crack.
A very important advantage of our modeling approach is that it allows us
to investigate the spatial structure of damage as well. Based on the
sample geometry and loading conditions illustrated in Fig.\
\ref{fig:sample}, the crack can be identified as 
a continuous region of broken beams starting from the bottom of the
interface. 
The high stress concentration
ahead of the crack tip and the quenched 
disorder of the local strength of beams give rise to the emergence of
a sequence of broken and intact beams followed by a continuous region
of intact elements. 
\begin{figure}
\begin{center}
\epsfig{bbllx=14,bblly=15,bburx=244,bbury=121,
file=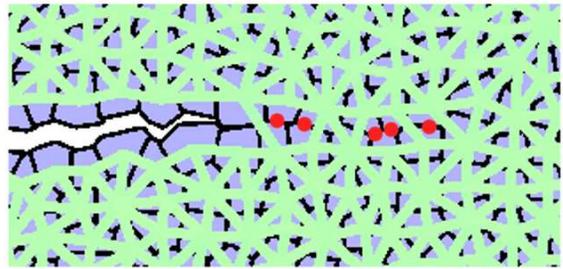, width=7.5cm}  
\caption{{\it (Color online)} Crack tip and the process zone in front
  of it. Red dots  
  indicate the positions of broken beams. The process zone is
  identified as a sequence of broken and intact beams starting at the
  crack tip and ending at the start of the continuous sequence of
  intact fibers. 
}
\label{fig:przone}
\end{center}
\end{figure}
Figure \ref{fig:przone} illustrates that in the
framework of our discrete element model the crack tip can be precisely
defined as the position of the first intact beam starting from the
bottom of the specimen. The sparse region of broken and intact beams,
extending from the crack tip to the last broken fiber, can be
identified as the fracture process zone (FPZ) whose dynamics has a
strong influence 
on the time evolution of the breaking process
\cite{bazant_fracture_1997}. We note that the 
extension of the process zone is also affected by the background
damage $D$ nucleated in an uncorrelated manner before the crack
starts (see Fig.\ \ref{fig:damratio}). The higher amount of background
damage in the bending limit 
results in a larger extension of the process zone than for the case of
stretching dominated breaking. 
\subsection{Dynamics of the process zone}
\begin{figure} [h]
\begin{center}
\epsfig{bbllx=1,bblly=507,bburx=341,bbury=792,
file=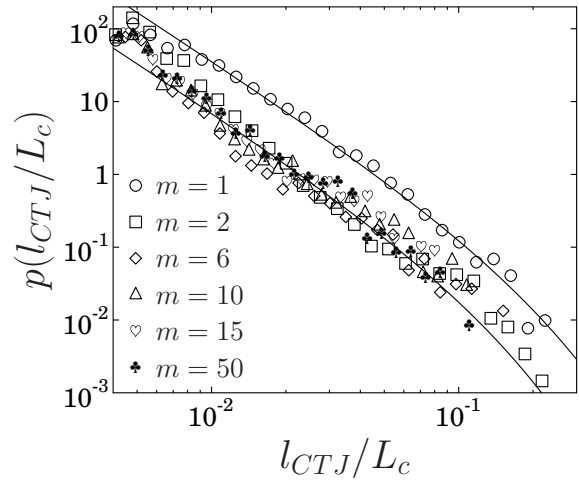, width=8.8cm}  
\caption{Crack tip jump length distributions for the absolute
  stretching limit. The value of the exponent of the power law regime
  is $\gamma^s =2.2 \pm 0.1$, it does not depend on the amount of
  disorder.
}
\label{fig:ctj_s}
\end{center}
\end{figure}

The dynamics of the process zone strongly determines the advancement
of the crack. As the beams break, the extension of the process zone
changes in discrete steps: when a new breaking nucleates
inside the intact zone the FPZ extends by a length $l_{nucl}$ called
nucleation length. The process zone shrinks when the beam at the crack
tip breaks resulting in a jump of the crack tip (CTJ) with a distance
$l_{CTJ}$. 
\begin{figure} [h]
\begin{center}
\epsfig{bbllx=1,bblly=507,bburx=341,bbury=792,
file=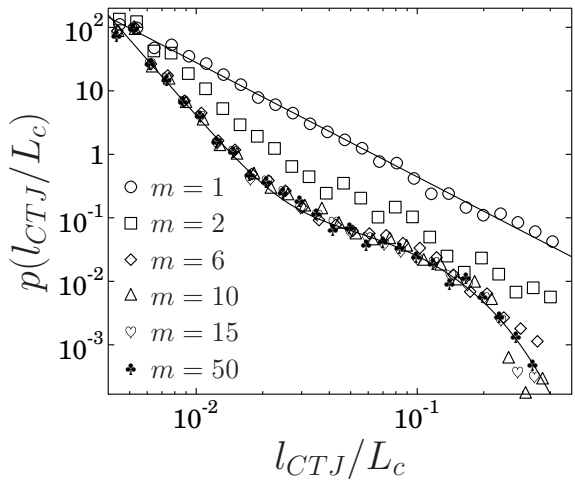, width=8.8cm}  
\caption{Crack tip jump length distributions for the absolute bending
  limit. The values of the exponent $\gamma^b$ of the fitted curves 
are $1.8$ and $4$. 
}
\label{fig:ctj_b}
\end{center}
\end{figure}
In order to characterize the dynamics of FPZ we investigate
the probability distribution of the nucleation length $p(l_{nucl})$
and of the length of crack tip jumps $p(l_{CTJ})$. We note that the
notion of crack tip jump length has also recently been introduced in the
framework of Quantized Fracture Mechanics
\cite{pugno_phil_mag_2004,pugno_ijf_2006_140,pugno_ijf_2006_141}. 
It can be observed in Fig. \ref{fig:ctj_b} that for low disorder (high
Weibull exponent $m$) in the bending limit
of breakings the distribution of crack tip jumps has a power law decay
in the regime of small $l_{CTJ}$ values which is complemented by an
exponential form for large $l_{CTJ}$
\begin{equation}
p(l_{CTJ})=a l_{CTJ}^{-\gamma} + c e^{-l_{CTJ}/b}.
\label{ctjdist}
\end{equation}
The additive coupling of the two terms of Eq.\ (\ref{ctjdist}) shows
that the small and large crack tip jumps are generated by different
mechanisms, i.e.\ the small ones are determined by the stress
concentration at the crack tip and by the resulting correlation of
local breakings. However, the Poisson-like behavior of very large
crack tip jumps captured by the second term of the equation originates
from the randomness of the initial jump-in of the crack at the onset
of crack propagation. 
The value of the exponent $\gamma^b$ changes from $\gamma^b=1.8 \pm 0.1$ to
$\gamma^b=4 \pm 0.2$ as the amount
of disorder decreases. In the stretching limit of breakings the sample
behaves in a less fragile way accumulating less background damage
before the onset of crack propagation (see also Fig.\
\ref{fig:damratio}). 
Hence, in  Fig.\ \ref{fig:ctj_s} we obtain a power law decay with an 
exponential cutoff but the additive exponential term does not
occur. Simulations showed that the value of the 
exponent does not depend 
on the amount of disorder $\gamma^s=2.2 \pm 0.1$ but it falls between
the low and high disorder limits of the bending counterpart.

\begin{figure} [h]
\begin{center}
\epsfig{bbllx=1,bblly=507,bburx=341,bbury=792,
file=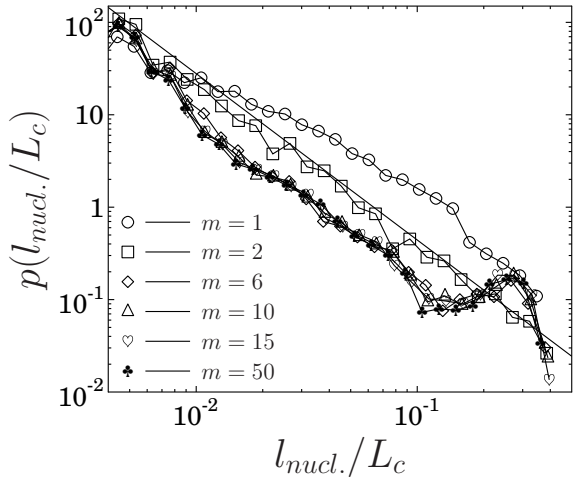, width=8.8cm}  
\caption{Nucleation length distributions for the absolute bending
  limit varying the Weibull exponent $m$. 
The amount of disorder does not have a relevant effect on
  the distribution. The exponent of the fitted power law is $\kappa^b=1.8$.
}
\label{fig:lnucl_b_40}
\end{center}
\end{figure}

The statistics of crack tip jumps was accumulated over the entire time
evolution of the bending process. In order to obtain information about
the nucleation length $l_{nucl}$, we analyze microcracks occurring
ahead of the crack tip at the time when the crack spans approximately
half of the cross section of the specimen. It can be observed in Fig.\
\ref{fig:lnucl_b_40} that in the bending limit the distribution 
$p(l_{nucl})$ hardly changes when the amount of disorder is varied. A 
power law form is obtained for low length values
\begin{eqnarray}
p(l_{nucl}) \sim l_{nucl}^{-\kappa}
\end{eqnarray}
with an exponent $\kappa^b=1.8 \pm 0.15$. When beam breaking is dominated
by tensile stresses the situation drastically changes: at high
disorder the nucleation distance can extend up to $40\%$ of the cross
section of the specimen as in the bending limit, i.e.\ very remote
beams may also break if 
they are weak enough. At low disorder, however, the nucleation of new
broken beams gets localized to the close vicinity of the crack tip
where the maximum of $l_{nucl}$ extends only up to $1-2\%$ of the cross
section. These results on the distance to new nucleations and on the
length of crack tip jumps clearly show that in the case of bending
dominated breaking varying the amount of disorder does not have a strong
effect on the spatial distribution of damage. The dynamics of the
process zone is mainly determined by the long range redistribution of
stresses arising from the bending distorsion. However, when breaking is
dominated by tensile deformation, disorder plays a crucial role in the
evolution of the fracture process zone, i.e.\ at low disorder the
process zone expands in a large number of small steps while shrinking
occurs in the form of a few larger jumps. When the disorder is high,
both shrinking and expanding steps can span a broad range. 

\subsection{Damage profile}
\begin{figure} [h]
\begin{center}
\epsfig{bbllx=1,bblly=507,bburx=341,bbury=792,
file=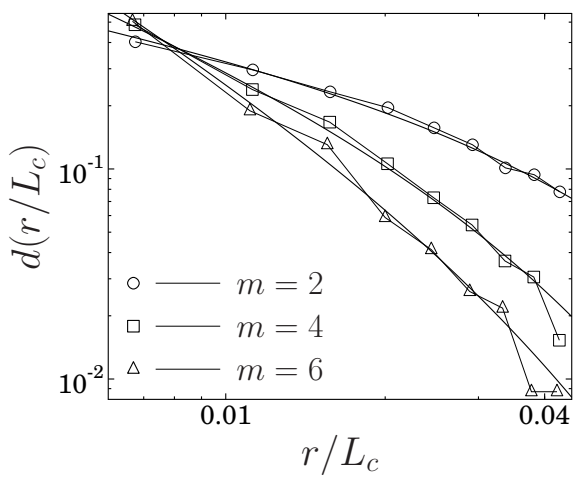, width=8.8cm}  
\caption{Damage profile for different values of the Weibull
  exponent $m$. The distance $r$ measured from the crack tip is
  normalized by the cross section $L_c$ of the specimen. 
The values of the fitting parameters are: $\rho=0.5$, 
$r_0=0.05$ $(m=2)$, $\rho=1.0$, $r_0=0.035$ $(m=4)$, $\rho=1.5$,
$r_0=0.033$ $(m=6)$. 
}
\label{fig:damprofile}
\end{center}
\end{figure}

To obtain quantitative estimates for the extension of the process zone,
we calculated the spatial distribution of damage in front of the crack
tip when the crack spans half of the specimen's cross section. Fig.\
\ref{fig:damprofile} presents the damage $d$, i.e.\ the 
probability of beam breaking as a function of the distance $r$ measured
from the crack tip for the stretching limit of the model. The
numerical results clearly demonstrate that 
larger Weibull exponents, i.e.\ higher degree of brittleness results in
smaller process zones. For all Weibull exponents the curves can be well 
described by a power law functional form with an exponential cutoff
\begin{eqnarray}
d(r) \sim r^{-\rho}\exp{(r/r_0)},
\label{eq:dr}
\end{eqnarray}
where the extension of the process zone can be characterized by the
length $r_0$. It can be observed in 
Fig.\ \ref{fig:damprofile} that both the exponent $\rho$ and the
characteristic length $r_0$ depend on the amount of disorder
$m$. Fitting the formula Eq.\ (\ref{eq:dr}) to the simulated data we
obtained the following parameter values: $\rho=0.5$, 
$r_0=0.05$ $(m=2)$, $\rho=1.0$, $r_0=0.035$ $(m=4)$, $\rho=1.5$,
$r_0=0.033$ $(m=6)$.

In order to obtain
an analytic understanding of the damage profile $d(r)$
we can start from the result of fracture mechanics
that in the vicinity of the crack tip the stress has a
power law decay 
\begin{eqnarray}
\sigma(r) = a r^{-\omega}.
\end{eqnarray}
For the exponent $\omega$ linear fracture mechanics predicts 
the value $\omega=1/2$ \cite{bazant_fracture_1997}, while fractal
cracks and plastic or hyper-elastic constitutive laws lead to
different values of $\omega$ \cite{pugno_efm_2008}.
Hence, the probability of beam breaking as a function of $r$ can be
estimated as 
\begin{eqnarray}
d(r) = P(\sigma(r)), 
\end{eqnarray}
where $P(x)$ is the cumulative distribution
function of the breaking thresholds. Since our
breaking thresholds are Weibull distributed,
we have $d(r) = 1 - e^{-(\sigma(r) /
  \lambda)^m} = 1 - e^{-(b r^{-\omega m})}$ where $b =
(a/\lambda)^m$. Restricting the calculation for small distance we
arrive at the form
\begin{eqnarray}
d(r) \approx b r^{-\omega m}. 
\label{eq:damprof}
\end{eqnarray}
The curves of Eq.\ (\ref{eq:dr}) in Fig.\ \ref{fig:damprofile} are
consistent with the 
analytical expression Eq.\ (\ref{eq:damprof}) for low values of $r$.
\begin{figure} [h]
\begin{center}
\epsfig{bbllx=1,bblly=507,bburx=341,bbury=792,
file=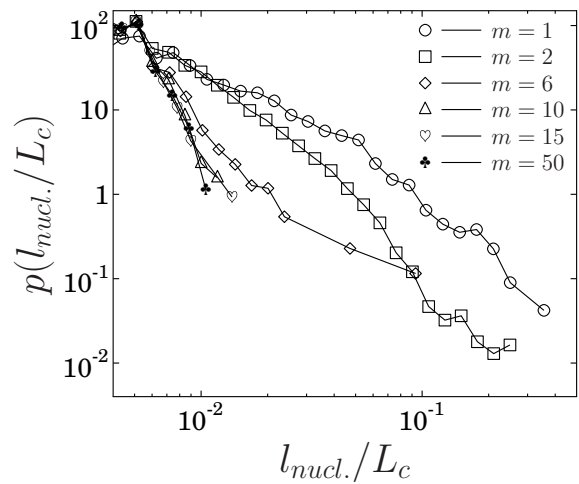, width=8.8cm}  
\caption{Nucleation length distributions for the absolute stretching
  limit. Decreasing disorder leads to localization of damage at the
  crack tip.
}
\label{fig:lnucl_s_40}
\end{center}
\end{figure}
Comparing the results to Eq.\ (\ref{eq:dr}) the exponent $\rho$
obtained by fitting the numerical data can be written as a product 
of the exponents of stress decay and disorder $\rho=\omega
m$. Substituting the numerical values of $\rho$ and the 
Weibull exponents $m$ the exponent $\omega$ describing the decay of
the stress field can be determined as $\omega\approx 0.25$ for all $m$
values. It is important to emphasize that this value of $\omega$ falls
quite close to the analytic result of $\omega=1/2$ of linear fracture
mechanics \cite{bazant_fracture_1997}, furthermore, the independence
of the numerically obtained 
$\omega$ from the Weibull exponent $m$ shows the consistency of the
results.  
The remarkable feature of the results is that the amount of
disorder can have a strong effect both on the shape and extension of the
process zone.

\section{Discussion} 
The emergence of crackling noise is a ubiquitous phenomenon of the
fracture of heterogeneous materials which can also be exploited to
monitor the time evolution of the fracture process. Theoretical
studies of crackling noise are 
usually based on stochastic fracture models such as fiber bundles
\cite{niccolini_crackling_2009,hidalgo_universality_2008,hidalgo_pre_2009} and
the fuse model \cite{alava_statistical_2006}. As a novel approach to
the problem, in the present 
paper we investigated the properties of crackling noise emerging
during the jerky propagation of a crack in three-point bending tests
using a discrete element modeling technique. Two limiting cases of
breakings were analyzed where stretching or bending dominates
the local breaking of beams. 
We proposed a numerical
technique to identify avalanches based on the temporal and spatial
correlation of micro-fractures. 

We showed that for quasi-brittle
materials the size of bursts and the waiting
times between consecutive events are characterized by power law
functional forms with an exponential cutoff. The numerical value of
the  exponents have a reasonable agreement with recent experimental
findings on crackling noise in three-point bending tests on concrete
specimens
\cite{niccolini_crackling_2009,niccolini_self-similarity_2009,pugno_jpcm_2006,pugno_small_2008,pugno_pre_2010_fiber,pugno_mst_2009,pugno_ieee_2009}.
This 
agreement also demonstrates the importance of spatial correlations of
consecutive microfractures in the emergence of crackling noise.

An important advantage of our DEM approach is that it provides direct
access to the spatial structure of damage. 
Simulations revealed that ahead of the crack tip a process zone develops
which is a sparse region 
of broken and intact elements. The fracture process zone proved to
play an important role in the advancement of the crack: on the one
hand the crack progresses by shrinking and expanding steps of the
zone, on the other hand, micro-cracks can shield  
the stress field around the crack tip which helps to stabilize the 
system. Recently the spatial structure of damage has been analyzed in
the framework of the fuse model
\cite{alava_role_2008,alava_size_2009}. Quasistatic loading
simulations were 
performed starting with a notch in the middle of the fuse lattice
analyzing the damage structure in the vicinity of the crack tip just
before macroscopic breakdown. It was found that the damage profile has
an exponential decay along the line of the crack and the characteristic
length scale of the exponential was suggested as the extension of the
process zone. Since linear fracture mechanics predicts a power law
decay of the stress to the background level ahead of the crack tip, the
authors argued that the cloud of microcracks shields the crack tip
giving rise to an exponential decay. In our system at short distances
a power law decay of the damage profile was obtained which is followed
by an exponential cutoff. We think the power law functional form
prevails in our system for the damage profile because microcrack
nucleation cannot occur in the two-dimensional plane but it is
restricted to a weak ``line'' in the sample which decreases the effect
of shielding.  This shielding, however, is responsible for the lower
exponent of the stress decay $\omega$ and for the exponential cutoff
of the damage profile. Computer simulation are complemented by
analytic calculations under simplifying conditions, which provided a
reasonable agreement with the numerical results.

\begin{acknowledgments}
The work is supported by T\'AMOP 4.2.1-08/1-2008-003 project. The
project is implemented through the New Hungary Development Plan,
co-financed by the European Social Fund and the European Regional
Development Fund. 
F.\ Kun
acknowledges the B\'olyai J\'anos fellowship of the Hungarian Academy of
Sciences.
\end{acknowledgments}

\bibliography{statphysfracture}

\end{document}